\documentclass[aps,prd]{revtex4}
\usepackage{graphicx}
\usepackage{latexsym}
\usepackage{amsmath}
\usepackage{amsfonts}
\usepackage{amssymb}
\usepackage{verbatim}

\newcommand{\be}{\begin{equation}}
\newcommand{\ee}{\end{equation}}
\newcommand{\bea}{\begin{eqnarray}}
\newcommand{\eea}{\end{eqnarray}}

\newcommand{\pa}{\partial}

\newcommand{\bb}{\bibitem}
\def\pls{\partial\!\!\!/}
\def\bb{\bibitem}

\def\ns{n\!\!\!/}

\def\bs{b\!\!\!/}

\def\bb{\bibitem}
\newcommand{\ben}{\begin{eqnarray}}
\newcommand{\een}{\end{eqnarray}}

\newcommand{\remark}[1]{}

\begin{document}

\title{Superluminal neutrinos from Lorentz-violating dimension-5 operators}
\author{C. A. G. Almeida$^{2}$, M. A. Anacleto$^{1}$, F. A. Brito$^{1}$ and E. Passos$^{1}$} 
\affiliation{$^{1}$Departamento de F\'\i sica,
Universidade Federal de Campina Grande, Caixa Postal 10071,
58109-970  Campina Grande, Para\'\i ba,
Brazil\\
$^{2}$Departamento de Ci\^encias Exatas, 
Universidade Federal da Para\'\i ba,
58297-000 Rio Tinto, Para\'\i ba, 
Brazil}


\begin{abstract}
We consider Lorentz- and CPT-violating dimension-5 operators to address the issue of superluminal neutrinos recently pointed out in OPERA experiments. We  assume these operators in the photon and neutrino sectors coupled to Lorentz-violating backgrounds in a preferred frame defined by a  time-like direction. We show that  such operators can produce a curve with OPERA's slope that fits OPERA, MINOS and supernova SN1987a data. 
\end{abstract}
\pacs{XX.XX, YY.YY} \maketitle



In this letter we consider Extended Myers-Pospelov dimension-5 operators in order to consider physics in a preferred frame with time-like direction in the presence of a Lorentz-violating background $n_\mu$ to adress the issue of the superluminal neutrinos detected in the OPERA's experiments very recently \cite{OPERA} --- for some recent theoretical developments see, for exemple \cite{OPERA2,Nojiri:2011ju,TL,CG}. We derive the dispersion relations associated to the
  effective Lagrangian for Dirac and Maxwell terms supplemented by dimension-5 operators given in the form
\bea\label{smf01}
{\cal L}_{eff}=\bar\psi(i\pls - m)\psi+\tilde{g}\bar\psi\,\ns\gamma_{5}\hat{D} \psi-\frac{1}{4}F_{\mu\nu}F^{\mu\nu}+g\epsilon^{\alpha\mu\lambda\rho}n_{\lambda}\hat{D}F_{\rho\alpha}A_{\mu}+...
\eea
where $(\cdots)$ means interacting terms,  $\tilde{g}=\eta/M$ and ${g}=\xi/M$ with $\eta,\xi$ dimensionless parameters. The dimension-5 operator in the electromagnetic sector is CPT - odd and even under charge conjugation, whereas the dimension-5 operator in the fermionic sector breaks CPT and is even under charge conjugation, $M$ is the mass where new physics such as Lorentz and CPT symmetry violation emerges and $\hat{D}$ is 
a derivative operator given by 
\bea\label{m51}
\hat{D}=(\partial\cdot n)^{2}-\partial^{2}n^{2}.
\eea
This is the  Myers-Pospelov operator \cite{MP} suitably extended in the present study to affect the superluminality behavior at large momenta. Note this reduces to the original operator for light-like backgrounds, i.e., for  $n^2=0$.
The fermionic sector with $\tilde{g}\,\hat{D}\ns\gamma_{5}\to -\bs\gamma_{5} $ reduces to CPT - odd extended QED by Colladay and Kostelecky \cite{kosteleck}, which has been recently considered in superluminal neutrino issues \cite{Alexandre:2011bu}. Similarly, the electromagnetic sector with $g\,n_{\mu}\,\hat{D}\to -\kappa_{\mu} $ reduces to the Maxwell and Carroll-Field-Jackiw model \cite{CFJ} . However, it was shown in  \cite{Adam:2001kx,Alfaro:2006dd} that this theory in time-like background does not produce superluminal velocities. So at this level one cannot expect to fit OPERA's data \cite{OPERA}. Thus, we should go to higher dimensional operators as in (\ref{smf01}). { Although such operators can produce superluminal velocities they comprise a Lorentz invariant Lagrangian except by the appearance of the Lorentz-violating background four vector $n^\mu$ \cite{MP,kosteleck}, which can also be understood as a vacuum expectation value (VEV) of a vector field \cite{Nojiri:2011ju, TL}. For the sake of simplicity, we choose time-like backgrounds, which is not necessarily the frame of the neutrinos. }

Let  us now derive the dispersion relation associated to extended Myers-Pospelov  dimension-5 operator in the fermionic sector.
The equation of motion takes the form
\bea\label{f09}
(i\pls - m - \tilde{g} \hat{D}\ns\gamma_{5})\psi=0.
\eea
After a straightforward algebra we find that the free continuous spectrum is governed
by the following dispersion relation
\bea\label{f10}
\big(k^{2}-m^{2}-\tilde{g}^{2}n^{2}P_{n}^{2}\big)^{2}-4 \tilde{g}^{2}P_{n}^{3}=0,
\eea
where
\bea\label{f11}
P_{n}=(n\cdot k)^{2}-n^{2}k^{2}.
\eea
For related issues in the original Myers-Pospelov model see, e.g., \cite{heyes, heyes2, mariz} --- see also \cite{king}.
Hence, settling a time-like direction in the Lorentz symmetry breaking background $n_{\mu}=(n_{0},0,0,0)$, we have
\bea\label{f12}
\big(\omega^{2}-|\vec{k}|^{2}-m^{2}-\tilde{g}^2n_{0}^{6}|\vec{k}|^{4}\big)^{2}-4\tilde{g}^{2}n_{0}^{6}|\vec{k}|^{6}=0,
\eea
whose solutions are 
\bea\label{f13}
\omega_{\pm}=\pm\sqrt{|\vec{k}|^{2}(1\pm \tilde{g} n_{0}^{3}|\vec{k}|)^{2}+m^{2}}.
\eea
We assume here the neutrino velocity is given by the group velocity determined from the fermionic dispersion relation (\ref{f13}), that is
\bea
v_\nu=\frac{\pa\omega_{+}}{\pa|\vec{k}|}=\frac{|\vec{k}|(1\pm \tilde{g}n_{0}^{3}|\vec{k}|)(1\pm 2\tilde{g}n_{0}^{3}|\vec{k}|)}{\sqrt{|\vec{k}|^{2}(1\pm \tilde{g} n_{0}^{3}|\vec{k}|)^{2}+m_\nu^{2}}}.
\eea
Now expanding in large momenta $|\vec{k}|^2\gg m_\nu^2$, but keeping $\tilde{g}n_{0}^{3}|\vec{k}|\ll1$ we find
\bea
v_\nu\simeq1\pm 2\tilde{g}n_{0}^{3}|\vec{k}|-\frac{m_\nu^2}{2|\vec{k}|^2}{(1-4\tilde{g}^2n_{0}^{6}|\vec{k}|^2)}.
\eea
If we mantain only linear terms in $\tilde{g}$ we find (for the ``plus sign'' sector)
\bea
{v_\nu-1}\sim 2\tilde{g}n_{0}^{3}|\vec{k}|-\frac{m_\nu^2}{2|\vec{k}|^2}.
\eea
Note that for massless or almost massless fermions as in the case of neutrinos we have possibility of superluminal velocities. We can indeed estimate the Lorentz violating coupling $\tilde{g}$ if we know the energy $E_\nu\sim |\vec{k}|$ and the mass $m_\nu$ of the superluminal  neutrino satisfying the bound
\bea
4\tilde{g}n_{0}^{3}|\vec{k}|^3>{m_\nu^2}.
\eea
As we shall see below, this is easily satisfied by OPERA's data  \cite{OPERA} where $\tilde{\alpha}\equiv\tilde{g}n_{0}^{3}\sim(1.7\times10^6\mbox{GeV})^{-1}.$

Let us now write the deviation of the neutrino velocity from the light speed (in the vacuum)  as a function of momenta according to the curve with slope $\tilde{\alpha}$
\bea\label{dev}
\frac{v_\nu-c}{c}\simeq2\,\tilde{\alpha}|\vec{k}|,
\eea
where we are choosing $c=1$.
For OPERA experiments ${v_\nu-1}\sim 10^{-5}$ with $E_\nu\sim|\vec{k}|\sim17$ GeV we find the slope
\bea\label{eq-En}
\tilde{\alpha}\sim\frac{1}{1.7\times10^{6}\rm{\ GeV}}\sim 10^{-22}{\rm\ m},
\eea
which sets our Lorentz symmetry breaking parameter $\tilde{\alpha}$ to test superluminality up to the energy $1000$ TeV, the energy scale present in primary cosmic rays originated in the explosion of massive stars \cite{Amenomori:2000br}. Now substituting (\ref{eq-En}) into (\ref{dev}) we get to the curve
\bea
\frac{v_\nu-c}{c}\sim 10^{-5}\left(\frac{|\vec{k}|}{17\rm{GeV}}\right).
\eea
Note this also agrees with MINOS experiment \cite{Adamson:2007zzb} with energy $E_\nu\sim|\vec{k}|\sim3$ GeV. To compare OPERA with supernova SN1987a measurements \cite{SN1987A} we use $E_\nu\sim|\vec{k}|\sim10$ MeV to obtain
\bea
\frac{v_\nu-c}{c}\sim 10^{-5}\left(\frac{10^{-3}}{1.7}\right),
\eea
that is consistent with the SN1987a bound $|v-c|/c<2\times10^{-9}$ \cite{SN1987A}.  We conclude that the above constraints on the deviation from the light speed  in superluminal neutrinos obey a {\it linear curve fitting} as a function of the energy with OPERA's slope $\tilde{\alpha}\sim 0.5\times 10^{-6}\rm{\ GeV}^{-1}$. This has also been noticed in other recent investigations \cite{Anacleto:2011bv,Tamburini:2011ic,Ellis:2011uk}.

Similarly we can find the dispersion relation for the electromagnetic sector through its corresponding equation of motion
\bea\label{m6}
\big(\pa^{2}\eta^{\mu\nu}-\pa^{\mu}\pa^{\nu}-2g\hat{D}\varepsilon^{\nu\mu\lambda\rho}n_{\lambda}\pa_{\rho}\big)A_{\nu}=0.
\eea
The photon dispersion relation in the time-like Lorentz-violating background is then given by
\bea\label{m8-2}
(\omega^{2}-|\vec{k}|^{2})^{2}-4g^{2}n_0^{6}|\vec{k}|^{6}=0.
\eea
Solving this equation we obtain the following solutions
\bea\label{m9}
\omega_{\pm}(\vec{k})=\pm|\vec{k}|\sqrt{1 \pm 2g n_{0}^{3}|\vec{k}|}.
\eea
The group velocity here determines the photon velocity 
\bea\label{m10}
v_\gamma=\frac{\pa\omega_{+}}{\pa|\vec{k}|}=\frac{1\pm 3g n^{3}_{0}|\vec{k}|}{\sqrt{1\pm 2g n^{3}_{0}|\vec{k}|}}.
\eea
For large momenta but keeping $|2g n^{3}_{0}\vec{k}|\ll1$ we find (for the ``plus sign" sector)
\bea\label{vg-photon}
v_\gamma-1\simeq 2g n^{3}_{0}|\vec{k}|.
\eea
Let us now use the group velocity for the photon given in (\ref{vg-photon}) to write the deviation of the neutrino velocity from the light speed (in the presence of the Lorentz-violating background)  as a function of momenta (up to linear terms)
\bea\label{dev-photon-neutrino}
\frac{v_\nu-v_\gamma}{v_\gamma}\simeq2\,(\tilde{\alpha}-\alpha)|\vec{k}|.
\eea
Recall that $\tilde{\alpha}\equiv\tilde{g}n_{0}^{3}$,  ${\alpha}\equiv{g}n_{0}^{3}$, $\tilde{g}=\eta/M$ and ${g}=\xi/M$, where we shall assume $n_0=1$ from now on.  The expected superluminality of the neutrino (with mass $m_\nu\sim0.2$ eV and energy $E_\nu\sim|\vec{k}|\sim$ 17 GeV) is  negligible
\bea\label{dev-photon-neutrino-exp}
\frac{v_\nu-v_\gamma}{v_\gamma}\simeq10^{-21}.
\eea
By using (\ref{dev-photon-neutrino}) allows us to set a bound to the couplings $\eta$ and $\xi$ that measure the strength of the time-like Lorentz-violating background acting into the neutrino and photon sectors. Thus, at the Planck scale $M\sim 10^{19}$ GeV it follows that
\bea\label{couplings}
\eta-\xi\simeq\frac{10^{-21}M}{2\times17\mbox{ GeV}}\sim 2.5\times10^{-4}.
\eea
The bound for $\eta$ can be found through the bound $\xi\sim 10^{-6}$ for photons as follows
\bea\label{couplings2}
\frac{\eta}{\xi}\simeq1+\frac{2.5\times10^{-4}}{\xi}\sim 10^2,
\eea
which corresponds to $\eta\sim10^{-4}$ for neutrinos. This means that at the Planck scale the neutrino interacts with the Lorentz violating background about 100 times as much as the photon interacts. Up to one order of magnitude these bounds approach those derived from renormalization group equations \cite{MP} --- see also \cite{casana} for other bounds via dimension-4 operators in purely space-like backgrounds.

If the couplings $\eta$ and $\xi$ change with energy keeping the difference $\eta-\xi$ then we consider it as the slope of the curve
\bea\label{dev-photon-neutrino-curve}
\delta v_\nu=\frac{v_\nu-v_\gamma}{v_\gamma}\simeq2\,(\eta-\xi)\frac{|\vec{k}|}{M}\sim 5\times 10^{-4}\frac{|\vec{k}|}{M}.
\eea
It is interesting to note that if we bring the Planck scale to the TeV scale, i.e., $M=M_{\rm Earth}\sim 1$ TeV and $|\vec{k}|\sim$ 17 GeV one finds the OPERA's result, i.e., about  $\delta v_\nu\sim10^{-5}$, for the deviation of the neutrino velocity from the light speed. { Furthermore, for the scale $M=M_{\rm Astro}>1000$ TeV it is sufficient to avoid astrophysical constraints since one can readily find $\delta v_\nu <10^{-9}$. This is in accord with \cite{TL} since in the latter case the  strength of the Lorentz- violating background $\tilde{\alpha}_{\rm Astro}\sim1/ M_{\rm Astro}$ is about $10^4$ times smaller than $\tilde{\alpha}_{\rm Earth}\sim 1/M_{\rm Earth}$  in the former case. One should note that the formula (\ref{dev-photon-neutrino-curve}) can also be readily applied to {\it electrons}. Finally, noticed that for photons one makes use of equation (\ref{vg-photon}) and $g\sim\tilde{g}\,  (\xi\sim\eta)$ on  the Earth --- this is consistent with (\ref{couplings}) --- to find that $\delta v_\gamma\sim 10^{-5}$ at $|\vec{k}|\sim$ 17 GeV, a result as also found in \cite{TL}.}

 To bring the Planck scale to lower scales one should consider extra-dimensions.  In the five-dimensional Randall-Sundrum scenario \cite{RS} one finds $M_5\approx M_{Planck}$ for the warp factor $e^{kL}$ about $10^{15}$ and $M\equiv M_5 e^{-kL}\sim$1TeV on the TeV (i.e., IR or Standard Model) brane.

{ Our Lagrangian is Lorentz invariant except by the appearance of the Lorentz-violating background four vector $n^\mu$. This is in the same
class of the theories \cite{MP,kosteleck,Nojiri:2011ju,TL}. As such, we can always choose a rest frame for the neutrino to study its decay to other particles. Because of the Lorentz invariance the energy-momentum is localy conserved and decay of highly energetic neutrinos into neutrinos and other particles with lower energies is forbidden \cite{Nojiri:2011ju}.  This should be enough to evade the Cohen-Glashow bound \cite{CG}. Furthermore, we can fine
tune $\xi$ and $\eta$ in order to make the Lorenz-violating background much stronger on the Earth than interstellar
scale to avoid all the astrophysical constraints on Lorentz violation and Cohen-Glashow bound as well discussed in \cite{TL}.}







{\acknowledgments} 

We would like to thank CNPq, CAPES, PNPD/PROCAD -
CAPES for partial financial support.


\end{document}